\documentclass[lettersize,journal]{IEEEtran}
\usepackage{amsmath,amsfonts}
\usepackage{algorithmic}
\usepackage{algorithm}
\usepackage{array}
\usepackage{amssymb}
\usepackage{textcomp}
\usepackage{stfloats}
\usepackage{url}
\usepackage{verbatim}
\usepackage{graphicx}
\usepackage{cite}
\usepackage{booktabs}
\usepackage{enumitem}
\usepackage{subfigure}
\usepackage{svg}
\usepackage[colorlinks,linkcolor=blue,citecolor=blue]{hyperref}
\usepackage{xcolor}
\usepackage{xpatch}
\makeatletter
\ExplSyntaxOn

\cs_new:Npn \bibColoredItems #1#2
  {
    \clist_map_inline:nn {#2} { \cs_new:cpn {bib@colored@##1} {#1} } 
  }
\ExplSyntaxOff

\newcommand\bib@setcolor[1]{%
  \ifcsname bib@colored@#1\endcsname
    \expanded{\noexpand\color{\csname bib@colored@#1\endcsname}}%
  \else
    \normalcolor
  \fi
}
\IfPackageLoadedTF{hyperref}{\@tempswatrue}{\@tempswafalse}
\if@tempswa
  \xpatchcmd\@bibitem {\H@item}{\bib@setcolor{#1}\H@item}{}{\PatchFailed}
  \xpatchcmd\@lbibitem{\H@item}{\bib@setcolor{#2}\H@item}{}{\PatchFailed}
\else
  \xpatchcmd\@bibitem {\item}  {\bib@setcolor{#1}\item}  {}{\PatchFailed}
  \xpatchcmd\@lbibitem{\item}  {\bib@setcolor{#2}\item}  {}{\PatchFailed}
\fi
\makeatother

\begin{document}

\title{$L^2$FMamba: Lightweight Light Field Image Super-Resolution with State Space Model}

\author{Zeqiang Wei\IEEEauthorrefmark{1}, Kai Jin\IEEEauthorrefmark{1}, Zeyi Hou, Kuan Song, Xiuzhuang Zhou, ~\IEEEmembership{Member,~IEEE}
\thanks{\hspace{-1em}\IEEEauthorrefmark{1} Equal contribution. Corresponding author: Xiuzhuang Zhou.}
\thanks{\hspace{-1em}Zeqiang Wei, Zeyi Hou, and Xiuzhuang Zhou are with the School of Artificial Intelligence, Beijing University of Posts and Telecommunications, Beijing 100088, China (email: weizeqiang@bupt.edu.cn, houziye@bupt.edu.cn, xiuzhuang.zhou@bupt.edu.cn).
Kai Jin is with the Bigo Technology Pte. Ltd., Beijing 100020, China (email: jinkai@bigo.sg). Kuan Song is with the Explorer Global (Suzhou) Artificial Intelligence Technology Co., Ltd., Suzhou, Jiangsu 215123, China (email: songkuan@explorer.global).}
\thanks{\hspace{-1em}This work was supported by {\color{black}{the National Natural Science Foundation of China under Grant 61972046,}} and additionally supported by BUPT Excellent Ph.D. Students Foundation under grant CX2021120.}
}

\maketitle

\begin{abstract}
Transformers bring significantly improved performance to the light field image super-resolution task due to their long-range dependency modeling capability.
However, the inherently high computational complexity of their core self-attention mechanism has increasingly hindered their advancement in this task.
To address this issue, we first introduce the LF-VSSM block, a novel module inspired by progressive feature extraction, to efficiently capture critical long-range spatial-angular dependencies in light field images. LF-VSSM successively extracts spatial features within sub-aperture images, spatial-angular features between sub-aperture images, and spatial-angular features between light field image pixels. On this basis, we propose a lightweight network, $L^2$FMamba {\color{black}{(Lightweight Light Field Mamba)}}, which integrates the LF-VSSM block to leverage light field features for super-resolution tasks while overcoming the computational challenges of Transformer-based approaches.
Extensive experiments on multiple light field datasets demonstrate that our method reduces the number of parameters and complexity while achieving superior super-resolution performance with faster inference speed. 
\end{abstract}


\section{Introduction}
\IEEEPARstart{U}{nlike} traditional 2D imaging that only records pixel intensity (i.e., spatial information), light field (LF) imaging captures the intensity and directional information of light rays (i.e., angular information), allowing for the precise depiction of light distribution in 3D space.
This grants LF images a broader range of functionalities in post-processing, such as refocusing \cite{jayaweera2020multi}, depth estimation \cite{zhang2016light}, and stereoscopic display \cite{hodges2020stereoscopic}. 
As a result, the flexibility of image editing and computational photography is enhanced, and new avenues are opened for research and applications across various domains.
The inherent hardware design of LF cameras creates a trade-off between angular and spatial resolutions, often resulting in sub-aperture images (SAIs) with low spatial resolution.
This limitation significantly restricts the practical applications of LF images.
Accordingly, the light field image super-resolution (LFSR) task has been proposed and extensively studied. 

Effectively capturing the inherent long-range spatial-angular contextual dependencies in LF images is key to solving the LFSR task.
Although scholars have attempted to use convolution-based methods \cite{wang2022disentangling, van2023light} to model and extract various spatial-angular features from different LF image representations, including sub-aperture image array (SAIs), macro-pixel images (MacPI), and epipolar plane images (EPI), they are constrained by the narrow receptive fields of convolution. This limitation hinders the full utilization of long-range spatial-angular features.
In recent years, numerous researchers have applied Transformer-based methods to tackle the LFSR task.
Transformer-based methods \cite{wang2022detail, liang2022light, Liang_2023_ICCV, cong2023lfdet, hu2024beyond} leverage the attention mechanism to effectively model long-range dependencies and extract spatial-angular features.
However, their quadratic complexity in terms of computation and memory significantly bottleneck their development.

Given the challenges posed by Transformer-based methods, particularly their high computational complexity, recent research has explored alternative approaches, and the state space model (SSM) \cite{chen2013state} has emerged as a strong competitor. Among these, Mamba \cite{gu2023mamba} and its variants have shown performance comparable to or even surpassing Transformers in tasks like computer vision and natural language processing. 
Inspired by these advances, we develop a Light Field Vision State Space Model (LF-VSSM) block to efficiently capture critical long-range spatial-angular dependencies in LFSR tasks through a progressive feature extraction strategy \cite{chen2023progressive, Liu_2024_CVPR}.
LF-VSSM successively extracts spatial features within sub-images, spatial-angular features between sub-images, and spatial-angular features between light field image pixels.
Consequently, it enhances the modeling of complex spatial-angular relationships, thereby improving LFSR performance. 
{\color{black}{The LF-VSSM module can replace the CNN or Transformer feature extraction components in existing LFSR networks, thereby enabling seamless integration with current architectures.}}
{\color{black}{Based on this module, we propose a novel lightweight network, $L^2$FMamba (Lightweight Light Field Mamba), specifically designed for light field image data. This network aims to improve reconstruction quality while reducing computational complexity, featuring fewer parameters, lower computational load, and faster inference speed, which in turn significantly enhances the overall performance {\color{black}{on}} the LFSR task.}}
The proposed method is validated on several widely used LFSR datasets, and the experimental results convincingly demonstrate its effectiveness.

In summary, the key contributions are as follows:

\begin{itemize}

    \item We introduce a novel LF-VSSM block that utilizes a progressive spatial-angular feature extraction strategy, significantly enhancing LFSR performance by effectively capturing long-range contextual dependencies within LF images.
        
    \item We propose a novel lightweight network, $L^2$FMamba, for LFSR tasks, significantly reducing the number of parameters and computational cost.


    \item We conduct extensive experiments on five commonly used LF datasets. Our method achieves the current state-of-the-art (SOTA) performance on 2x and 4x LFSR tasks. Ablation studies further validate the effectiveness, efficiency, and scalability of the proposed network.
    
\end{itemize}

\section{Related Work}

Unlike single-image super-resolution (SISR) tasks, LF images typically consist of multiple SAIs with interrelated and complementary angular and spatial information.
Therefore, effectively leveraging this complementary information is often crucial for determining LFSR performance.

\subsection{CNN-based Light Field Super-Resolution}
Since LFCNN \cite{yoon2015learning} first introduced convolutional neural networks (CNN) into LFSR, various improvements based on CNN have emerged continuously, and modeling methods have been constantly evolving, making it the primary approach for LFSR tasks.
With LF-InterNet \cite{wang2020spatial}, two CNN feature extractors were designed to decouple and extract spatial and angular features, and the spatial-angular interaction mechanism was proposed to gradually integrate the extracted LF image features, thereby significantly enhancing LFSR performance.
DistgSSR \cite{wang2022disentangling} introduced the disentangling mechanism, extracting spatial, angular, and EPI information separately from various representations of LF data.
However, these methods are limited by the narrow receptive fields of convolutions, hindering their ability to effectively model long-range spatial-angular features in LF images, which ultimately affects the performance of the LFSR task.

\subsection{Transformer-based Light Field Super-Resolution}
In recent years, Transformers have achieved tremendous success and rapid development in computer vision due to its self-attention mechanism. 
Compared to CNN, this self-attention mechanism enables the Transformer to capture global relationships and long-range dependencies in images effectively. 
In DPT \cite{wang2022detail}, a spatial-angular locally-enhanced self-attention layer was designed to extract non-local contextual information from multiple views and preserve image details for each single view, thus better characterizing the geometric structure of LF images.
A new paradigm was proposed in LFT \cite{liang2022light}, namely angular and spatial Transformers, aimed at integrating angular and spatial information in LF. With a smaller model size and lower computational cost, it achieved superior SR performance.
EPIT \cite{Liang_2023_ICCV} employed Transformers to model horizontal and vertical EPIs, enabling the learning of non-local spatial-angular correlations, thereby enhancing the robustness of SR under large-disparity conditions.
LF-DET \cite{cong2023lfdet} relied on multi-scale angular modeling to address SR of scenes under different disparity ranges and reduced the computational cost required for global spatial feature modeling of SAIs by introducing subsampling spatial modeling.
{\color{black}{M2MT \cite{hu2024beyond} reveals the subspace isolation limitation caused by the decomposition and proposes a many-to-many transformer to achieve cross-subspace global interaction, thereby significantly enhancing LFSR performance through non-local context modeling.}}
Although these works attempt to address the high computational complexity of Transformers, they still suffer from the inherent limitations of the attention mechanism, as convolution is limited by its local receptive field. Therefore, an efficient alternative solution is urgently needed.

\subsection{Mamba-based Light Field Super-Resolution}
Mamba \cite{gu2023mamba} features linear-time sequence modeling with a selective state space model (SSM) \cite{chen2013state} layer and a hardware-friendly design, effectively addressing the computational challenges that Transformers face when processing long sequences. This allows for more efficient capture of long-range dependencies while significantly accelerating inference.
Since Vision Mamba \cite{zhu2024vision} and Vmamba \cite{liu2024vmamba} successfully introduced Mamba into computer vision, this model has found broad applications in tasks such as detection, segmentation, image restoration, point cloud processing, and 3D reconstruction, leading to a significant increase in related publications.
MambaIR \cite{guo2024mambair} was the first to apply this model to image restoration, introducing an enhanced Residual State Space module that effectively leverages local information and reduces redundancy.
This approach has outperformed strong CNN and Transformer baselines in various image restoration tasks, underscoring the powerful potential of the Mamba model in this domain.
DVMSR \cite{lei2024dvmsr} further compressed the parameters of the Mamba model through a feature distillation strategy, proposing a lightweight super-resolution model delivering comparable performance with fewer parameters.
While Mamba-based super-resolution methods have demonstrated excellent performance in SISR tasks, their direct application to LF images often yields suboptimal results as they neglect the unique spatial and angular characteristics of such images.

{\color{black}{LFMamba \cite{xia2024lfmamba} was the first to introduce Mamba into the LFSR task by applying SSM to various two-dimensional slices of the LF image, thereby effectively extracting spatial context, angular, and structural information. 
 In contrast, MLFSR \cite{Gao_2024_ACCV} proposed a bidirectional subspace scanning scheme to efficiently model the complex four-dimensional correlations inherent in LF images.
Both approaches employ vertical and horizontal Epipolar Plane Images (EPIs) to model angular relationships. 
However, EPI-based methods rely solely on information from a single epipolar line, which imposes significant locality constraints{\color{black}{;}} they neglect the intricate structural relationships between different epipolar lines and thus struggle to capture global dependencies across these lines, ultimately limiting the integration of global contextual information. 
In contrast, our method directly performs angular relationship learning on SAIs and MacPI, with SAIs used to model inter-image angular correlations and MacPI dedicated to capturing angular variations at the pixel level.
This approach not only effectively overcomes the shortcomings of EPI-based methods in capturing cross-epipolar information, {\color{black}{but}} also leverages the long-range dependency modeling capability of the Mamba framework by comprehensively integrating global contextual information across the entire LF images.
This further enhances LFSR performance while reducing parameters and computational complexity.
}}

\begin{figure*}[!t]
    \centering
    \includegraphics[width=18cm]{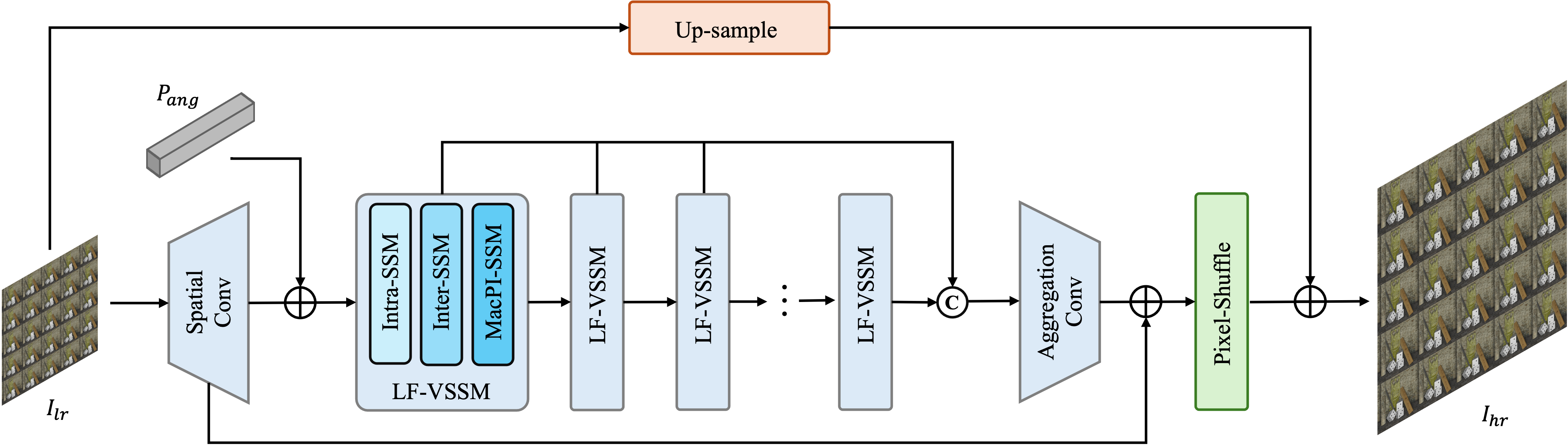}
    \caption{The architecture of the proposed $L^2$FMamba. \textcircled{c} represents feature concatenation, and $\bigoplus$ represents feature addition.}
    \label{fig:homepage}
\end{figure*}

\section{Method}
This section first provides an overview of the network architecture of $L^2$FMamba, followed by a detailed explanation of the design rationale behind the LF-VSSM block.

\subsection{Overview}

Start with a low-resolution LF image $I_{lr} \in \mathbb{R}^{(U \times H) \times (V \times W) \times 1}$, where $U \times V$ denotes the angular resolution, $H \times W$ represents the spatial resolution of the SAI, and 1 refers to the $Y$ channel in the $YCbCr$ color space.
Based on \cite{liang2022light, Liang_2023_ICCV, cong2023lfdet}, an initial spatial convolution $SpaConv$ is designed to obtain the initial LF features $F_{init} \in \mathbb{R}^{(U \times V) \times H \times W \times C}$, where $C$ represents the feature dimension, which is set to a default value of 64.
To extract angular information more effectively, a learnable angular positional encoding is designed, $P_{ang} \in \mathbb{R}^{(U \times V) \times 1 \times 1 \times C}$. This encoding is then added to $F_{init}$ to produce angularly enriched features, denoted as $F_{init}^{ang}$.
While spatial positional encoding is commonly used in Transformer-based methods, our design incorporates multiple selective scan directions in SSM, thus providing the model with spatial position awareness, as validated by our ablation experiments.
Moreover, fixed-length encoding may have adaptability issues across different resolutions, potentially compromising the robustness of the model. Therefore, we chose not to implement any spatial positional encoding.
Subsequently, $F_{init}^{ang}$ through $K$ sequentially stacked LF-VSSM blocks, resulting in $\{Out_i \in \mathbb{R}^{(U \times V) \times H \times W \times C} | i \in [1, K]\}$, where $K$ is set to a default of 4.
The above outputs are concatenated along the channel dimension and passed through $AggConv$ to obtain the aggregated feature $F_{agg}$, where $AggConv$ is a simple single-layer convolution neural network with the kernel size of {\color{black}{$3\times3\times (K \times C) \times C$}}.
Following the typical setup of deep super-resolution tasks, we employed residual learning methods to capture subtle differences between high- and low-resolution images rather than aiming for a direct and exact mapping from low to high resolution.
The LF image residual $R_{hr} \in \mathbb{R}^{(U \times V) \times \alpha H \times \alpha W \times 1}$ is obtained through the pixel shuffle residual upsampling module, where $\alpha$ represents the ratio of super-resolution, which is set to 2 or 4 in this paper.
The final high-resolution LF image $I_{hr}$ is obtained through the up-sampling module of $I_{lr}$ and adding residual output $R_{hr}$.
Fig. \ref{fig:homepage} illustrates the network architecture and data pipeline of $L^2$FMamba.

\begin{figure*}[!t]

    \centering
    \subfigure[Intra-SSM]{
        \label{structure:intra-ssm}
        \includegraphics[width=5.4cm]{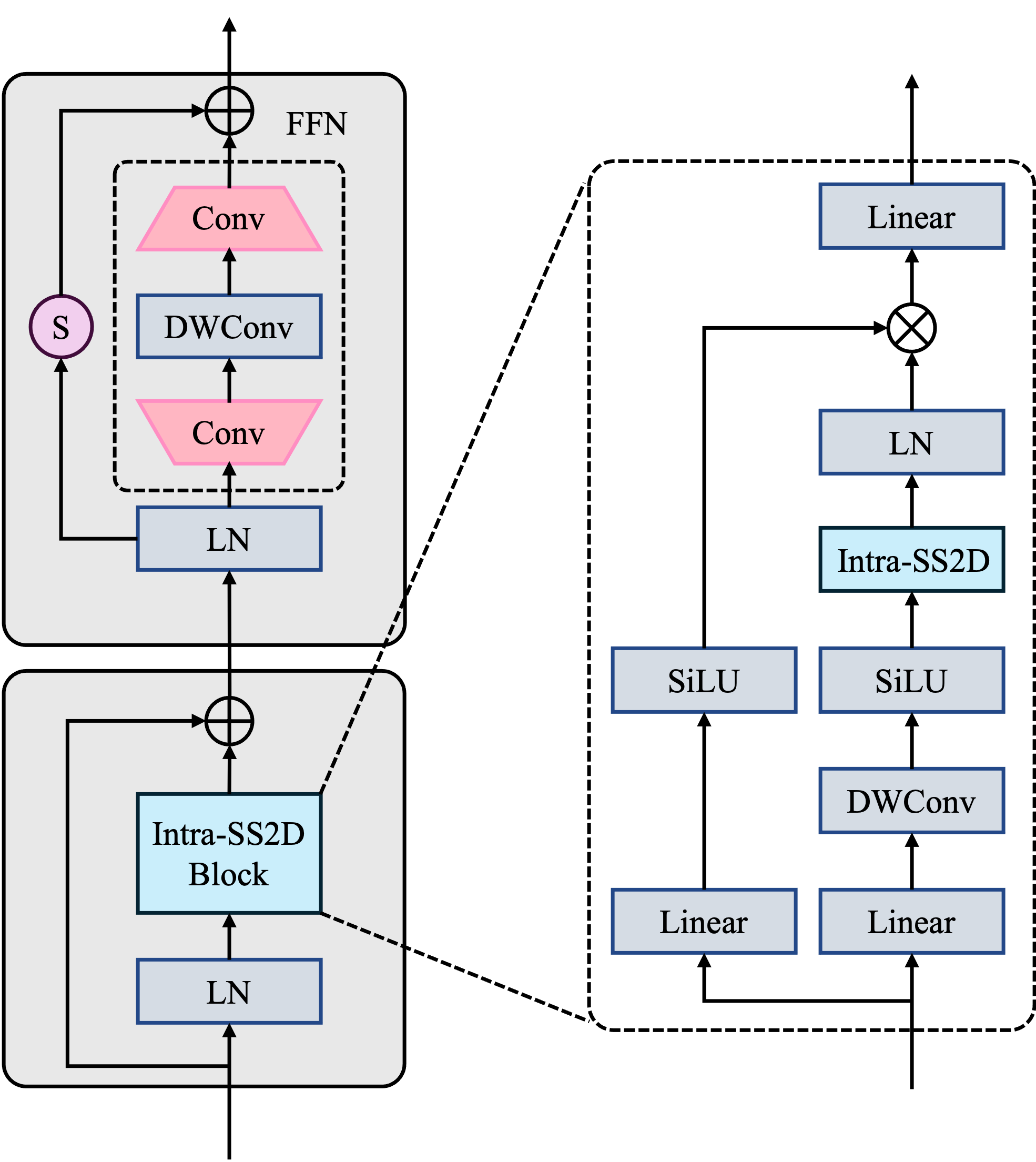}
    }
    \hspace{10pt}
    \subfigure[Inter-SSM]{
        \label{structure:inter-ssm}
        \includegraphics[width=5.4cm]{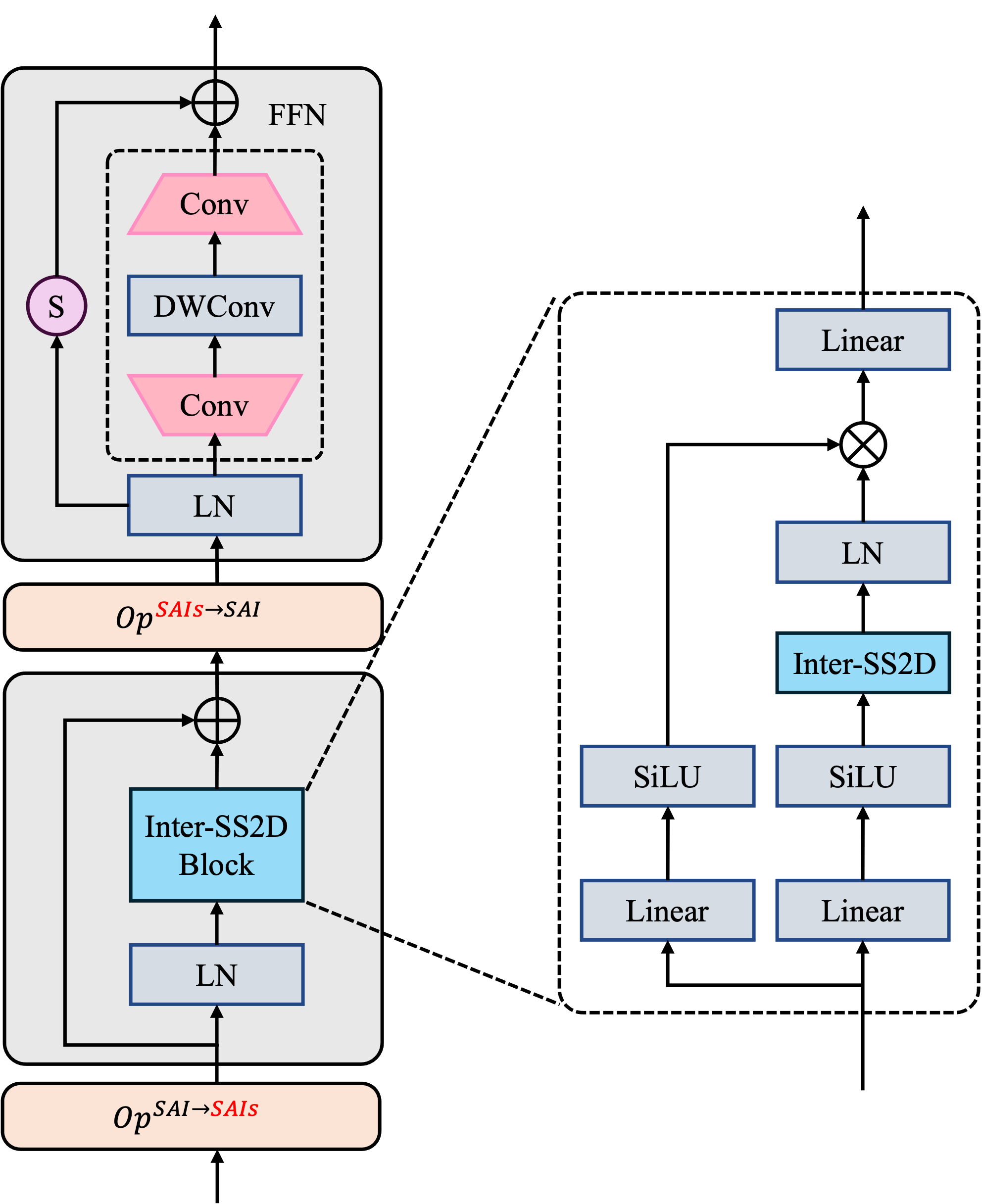}
    }
    \hspace{10pt}
    \subfigure[MacPI-SSM]{
        \label{structure:macpi-ssm}
        \includegraphics[width=5.4cm]{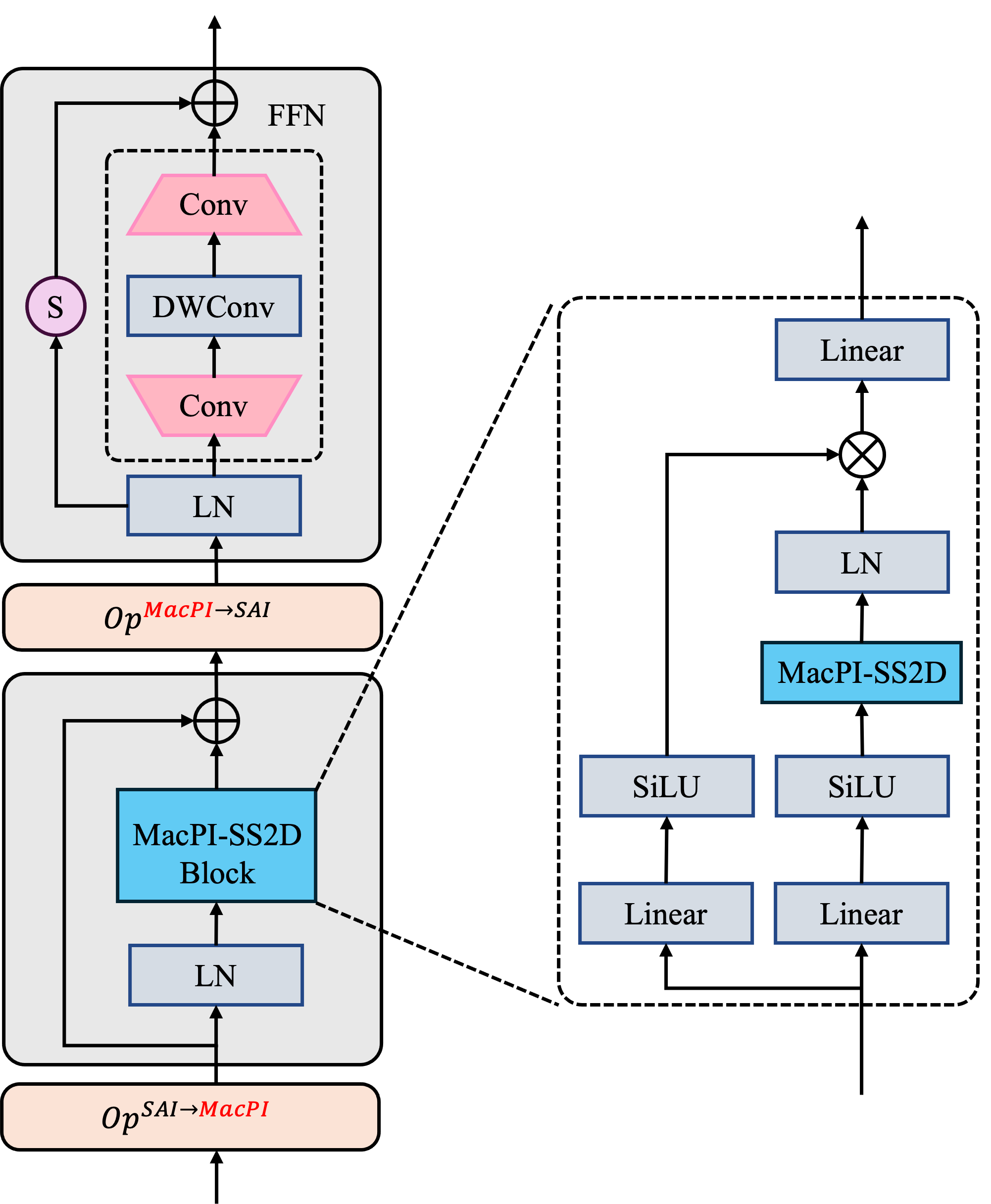}
    }
\caption{{\color{black}{Illustration {\color{black}{of}} the network structure of (a) Intra-SSM, (b) Inter-SSM, and (c) MacPI-SSM. \textcircled{s} represents the learnable adjustment factor $\lambda$.}}}
\label{SSM_Network}
\end{figure*}

\subsection{LF-VSSM Block}
The LF-VSSM block consists of three sequential sub-modules: {\color{black}{Intra-SSM}} for spatial feature extraction, {\color{black}{Inter-SSM}} for spatial-angular features between SAIs, and MacPI-SSM for inter-macro-pixel spatial-angular features extraction. These modules progressively integrate the angular features of the light field with the spatial features of the images, thereby enhancing the feature representation learning necessary for LFSR tasks.



\subsubsection{{\color{black}{Intra-SSM}}}
The {\color{black}{Intra-SSM}} module takes SAI features $In_{SAI} \in \mathbb{R}^{(U \times V) \times H \times W \times C}$ as input and focuses on spatial feature extraction, as shown in Fig. \ref{structure:intra-ssm}. 
Similar to conventional Transformer architectures, this module includes {\color{black}{Intra 2D-Selective-Scan (Intra-SS2D)}} and Feed-Forward Neural Network (FFN) layers. First, we normalize $In_{SAI}$ using layer normalization (LN) to stabilize training and improve convergence speed.
{\color{black}{Intra-SS2D}} is then employed to model global contextual relationships within each SAI, thus enhancing spatial feature extraction. This process, expressed as
\begin{equation}
    Z_{SAI} = \text{Intra-SS2D}(LN(In_{SAI})) + In_{SAI}
\end{equation}
is crucial for maintaining and propagating spatial information effectively through residual connections.

Since this module is dedicated to spatial feature extraction from SAIs without considering angular information, the images are treated as regular 2D ones. We apply the 2D selective scanning method from VMamba \cite{liu2024vmamba}, unfolding the 2D image into four 1D sequences in top-down, left-right, bottom-up, and right-left directions, as shown in Fig. \ref{direction:intra-ss2d}. After feature extraction, these sequences are merged to reconstruct the original 2D shape.

{\color{black}{Inspired by LocalViT \cite{li2021localvit}, we augmented the FFN module by incorporating a $3\times3$ depth-wise convolution (DWConv) layer. This layer leverages its local receptive field to better capture and integrate local spatial features within the SAI.}}
Additionally, a learnable adjustment factor $\lambda$ is applied at the residual connection of FFN to control the relative importance between global and local features:
\begin{equation}
    Out_{SAI} = \text{DWConv}(LN(Z_{SAI})) + \lambda \cdot Z_{SAI}
\end{equation}
This layer ensures that global and local spatial patterns are effectively captured, providing consistent improvements in spatial feature aggregation and representation across the model.

{\color{black}{It is worth noting that in the Intra-SS2D module, the DWConv layer is used to capture the local spatial structure of the input image, thereby extracting more fine-grained local features. Meanwhile, in the FFN module, the DWConv layer is applied after the global attention module has extracted features, further integrating and enhancing the aggregation of local information. Although both modules employ DWConv, they serve different purposes in capturing local details and consolidating local information, forming a complementary design.}}





\subsubsection{{\color{black}{Inter-SSM}}}

We restructure the {\color{black}{Intra-SSM}} outputs from the same LF image into $In_{SAIs} \in \mathbb{R}^{(U \times H) \times (V \times W) \times C}$. This serves as the input to {\color{black}{Intra-SSM}}. 
This module aims to extract spatial features within SAIs, which are sub-images of the same SAI image, while also capturing long-range angular dependencies across these multiple SAIs.
The {\color{black}{Inter-SSM}} network structure, as shown in Fig. \ref{structure:inter-ssm}, is similar to that of {\color{black}{Intra-SSM}} but with two key differences.
First, convolution operations applied to SAI features can disrupt the original spatial structure, introducing unnecessary noise at the edges where the SAIs connect. To mitigate this defect, we have removed the DWConv operation after the MLP.
Second, in the FFN layer, we introduce a transformation operator that converts SAI features back into SAI features, enabling effective integration of local spatial and global light field features.
The computation process of {\color{black}{Inter-SSM}} can be formalized as follows:
\begin{equation}
\begin{split}
    In_{SAIs} &= Op^{SAI \to SAIs}(Out_{SAI}) \\
    Z_{SAIs}  &= \text{Inter-SS2D}(LN(In_{SAIs})) + In_{SAIs} \\
    Z_{SAIs}  &= Op^{SAIs \to SAI}(Z_{SAIs}) \\
    Out_{SAIs}  &= \text{DWConv}(LN(Z_{SAI})) + \lambda \cdot Z_{SAIs}
\end{split}
\end{equation}

\begin{figure*}[!t]

    \centering
    \subfigure[Intra-SS2D]{
        \label{direction:intra-ss2d}
        \includegraphics[width=18cm]{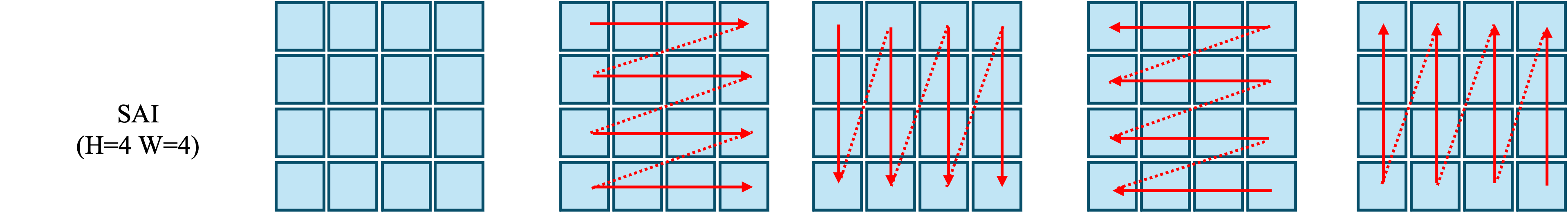}
    }
    \hspace{10pt}
    \subfigure[Inter-SS2D]{
        \label{direction:inter-ss2d}
        \includegraphics[width=18cm]{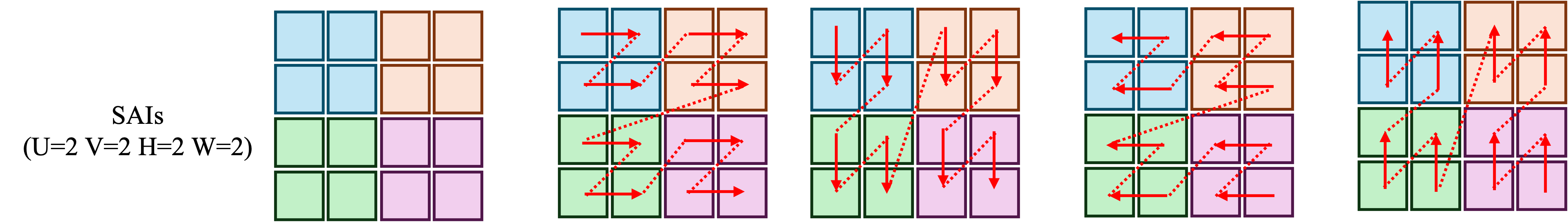}
    }
    \hspace{10pt}
    \subfigure[MacPI-SS2D]{
        \label{direction:macpi-ss2d}
        \includegraphics[width=18cm]{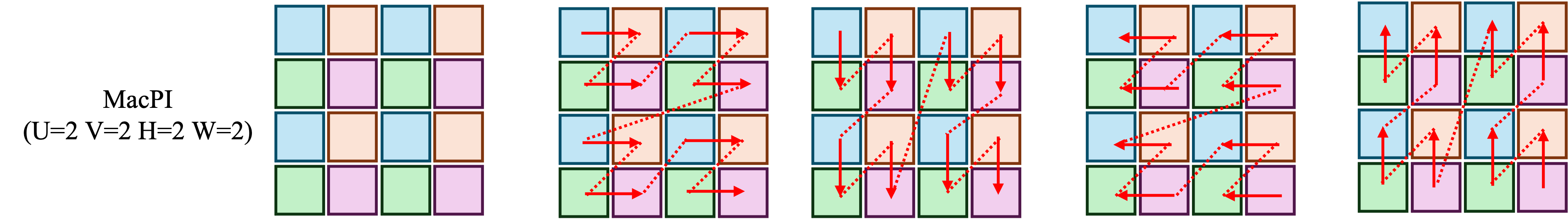}
    }
\caption{{\color{black}{Illustration of the selective scan directions of (a) Intra-SS2D, (b) Inter-SS2D, and (C) MacPI-SS2D. The first column represents the inputs of different LF data structures, where the same color indicates the same viewpoint, and different colors indicate different viewpoints. The last four columns represent four different scanning directions, with the red arrows indicating the scanning route and order.}}}
\label{SSM_Network}
\end{figure*}

 To adapt to the representation of SAIs in LF images, we adjusted the scanning directions in the {\color{black}{Inter-SS2D}} layer. Following a progressive feature extraction design, this module first prioritizes extracting spatial features within SAIs before capturing angular contextual relationships between SAIs.
We decompose the scanning directions into two components: spatial scanning directions within the image, $D_{spa}=\{d^1_{spa},d^2_{spa},d^3_{spa},d^4_{spa}\}$, and angular directions between images, $D_{ang}=\{d^1_{ang},d^2_{ang},d^3_{ang},d^4_{ang}\}$. Indices $1\text{-}4$ correspond to the top-down, left-right, bottom-up, and right-left directions. Spatial scanning is performed first, followed by angular scanning, forming a complete scanning path shown in Fig. \ref{direction:inter-ss2d}.

The window-based scanning method in LocalMamba \cite{yao2024spectralmambaefficientmambahyperspectral} is the most similar to our approach. When the window size is set to match the SAI size, the methods are identical in implementation. However, our method starts with scanning within the SAI image to better extract spatial-angular features, contrasting with LocalMamba, which focuses on image-bias induction for finer local feature extraction.

\subsubsection{MacPI-SSM}
We restructure the {\color{black}{Inter-SSM}} outputs into $In_{MacPI} \in \mathbb{R}^{(H \times U) \times (W \times V) \times C}$, which serves as the input to MacPI-SSM.
This module aims to extract long-range angular contextual dependencies between pixels in SAIs.
The MacPI-SSM network structure is essentially the same as that of the above {\color{black}{Inter-SSM}}, and the design principles are also similar, except that a transformation operator from MacPI to SAI is used instead of the corresponding operation in the FFN layer, as shown in Fig. \ref{structure:macpi-ssm}. The computation process of MacPI-SSM is as follows:
\begin{equation}
\begin{split}
    In_{MacPI} &= Op^{SAI \to MacPI}(Out_{SAIs}) \\
    Z_{MacPI}  &= \text{MacPI-SS2D}(LN(In_{MacPI})) + In_{MacPI} \\
    Z_{MacPI}  &= Op^{MacPI \to SAI}(Z_{MacPI}) \\
    Out_{MacPI}  &= \text{DWConv}(LN(Z_{MacPI})) + \lambda \cdot Z_{MacPI}
\end{split}
\end{equation}

In the MacPI-SS2D layer, we redesign the scanning method to accommodate the MacPI data format by introducing a two-part scanning method for spatial and angular information. 
Specifically, the angular scanning $D_{ang}$ is performed first within each macro-pixel, followed by spatial scanning $D_{spa}$ between macro-pixels. The directions indexed by $\{d^i_{ang}, d^i_{spa}\}$ form the complete scanning path, as detailed in Fig. \ref{direction:macpi-ss2d}.

\begin{table*}[!t]
    \centering
    \caption{PSNR/SSIM results compared with SOTA methods for 2$\times$ and 4$\times$ LFSR tasks. 
             The {\color{black}{best}} and the second best results are {\color{black}{respectively in bold and}} underlined.}
    \begin{tabular}{@{}cccccccc@{}}
        \toprule[2pt]
        Method                                  & Scale        & EPFL           & HCINew         & HCIold         & INRIA          & STFgantry      & Average        \\ 
        \midrule[1pt]  
        RCAN        \cite{zhang2018image}        & $\times$2    & 33.156/0.9635  & 35.022/0.9603  & 41.125/0.9875  & 35.036/0.9769  & 36.670/0.9831  & 36.202/0.9743  \\
        resLF       \cite{zhang2019residual}     & $\times$2    & 33.617/0.9706  & 36.685/0.9739  & 43.422/0.9932  & 35.395/0.9804  & 38.354/0.9904  & 37.495/0.9817  \\
        LFSSR       \cite{yeung2018light}        & $\times$2    & 33.671/0.9744  & 36.802/0.9749  & 43.811/0.9938  & 35.279/0.9832  & 37.944/0.9898  & 37.501/0.9832  \\ 
        LF-ATO      \cite{jin2020light}          & $\times$2    & 34.272/0.9757  & 37.244/0.9767  & 44.205/0.9942  & 36.171/0.9842  & 39.636/0.9929  & 38.306/0.9847  \\
        MEG-Net     \cite{zhang2021end}          & $\times$2    & 34.312/0.9773  & 37.424/0.9777  & 44.097/0.9942  & 36.103/0.9849  & 38.767/0.9915  & 38.141/0.9851  \\
        DistgSSR    \cite{wang2022disentangling} & $\times$2    & 34.809/0.9787  & 37.959/0.9796  & 44.943/0.9949  & 36.586/0.9859  & 40.404/0.9942  & 38.940/0.9867  \\
        LF-InterNet \cite{wang2020spatial}       & $\times$2    & 34.112/0.9760  & 37.170/0.9763  & 44.573/0.9946  & 35.829/0.9843  & 38.435/0.9909  & 38.024/0.9844  \\
        LF-IINet    \cite{liu2021intra}          & $\times$2    & 34.736/0.9773  & 37.768/0.9790  & 44.852/0.9948  & 36.564/0.9853  & 39.894/0.9936  & 38.763/0.9860  \\
        HLFSR-SSR   \cite{vinh2023-lfsr}         & $\times$2    & 35.310/0.9800  & \underline{38.317}/\underline{0.9807}  & 44.978/0.9950  & 37.060/\underline{0.9867}  & 40.849/0.9947  & 39.303/0.9874  \\
        LFSSR\_SAV  \cite{cheng2022spatial}      & $\times$2    & 34.616/0.9772  & 37.425/0.9776  & 44.216/0.9942  & 36.364/0.9849  & 38.689/0.9914  & 38.262/0.9851  \\
        DPT         \cite{wang2022detail}        & $\times$2    & 34.490/0.9758  & 37.355/0.9771  & 44.302/0.9943  & 36.409/0.9843  & 39.429/0.9926  & 38.397/0.9848  \\
        LFT         \cite{liang2022light}        & $\times$2    & 34.783/0.9776  & 37.766/0.9788  & 44.628/0.9947  & 36.539/0.9853  & 40.408/0.9941  & 38.825/0.9861  \\
        EPIT        \cite{Liang_2023_ICCV}       & $\times$2    & 34.826/0.9775  & 38.228/\textbf{0.9810}  & \textbf{45.075}/\underline{0.9949}  & 36.672/0.9853  & \textbf{42.166}/\textbf{0.9957}  & 39.393/0.9869  \\
        LF-DET      \cite{cong2023lfdet}         & $\times$2    & 35.262/0.9797  & 38.314/0.9807  & \underline{44.986}/\textbf{0.9950}  & 36.949/0.9864  & \underline{41.762}/0.9855  & \underline{39.455}/\underline{0.9874}  \\
        {\color{black}{MLFSR  \cite{Gao_2024_ACCV}}}       & {\color{black}{$\times$2}}    & {\color{black}{35.218/\underline{0.9801}}}  & {\color{black}{38.140/0.9803}}  & {\color{black}{44.904/0.9950}}  & {\color{black}{36.919/0.9865}}  & {\color{black}{40.975/0.9949}} & {\color{black}{39.231/0.9873}} \\
        {\color{black}{LFMamba  \cite{xia2024lfmamba}}}    & {\color{black}{$\times$2}}    & {\color{black}{\textbf{35.758}/\textbf{0.9824}}}  & {\color{black}{\textbf{38.368}/0.9801}}  & {\color{black}{44.985/0.9950}}  & {\color{black}{\underline{37.063}/\textbf{0.9876}}}  & {\color{black}{40.954/0.9948}} & {\color{black}{39.424/\textbf{0.9881}}} \\
        $L^2$FMamba (Ours)                       & $\times$2    & \underline{35.515}/0.9796  & 38.225/0.9803  & 44.953/0.9949  & \textbf{37.165}/0.9862  & 41.567/\underline{0.9952}  & \textbf{39.485}/0.9873  \\
        \midrule[1pt]  \midrule[1pt]  
        RCAN        \cite{zhang2018image}        & $\times$4    & 27.904/0.8863  & 29.694/0.8886  & 35.359/0.9548  & 29.800/0.9276  & 29.021/0.9131  & 30.355/0.9141  \\
        resLF       \cite{zhang2019residual}     & $\times$4    & 28.260/0.9035  & 30.723/0.9107  & 36.705/0.9682  & 30.338/0.9412  & 30.191/0.9372  & 31.243/0.9322  \\
        LFSSR       \cite{yeung2018light}        & $\times$4    & 28.596/0.9118  & 30.928/0.9145  & 36.907/0.9696  & 30.585/0.9467  & 30.570/0.9426  & 31.517/0.9370  \\ 
        LF-ATO      \cite{jin2020light}          & $\times$4    & 28.514/0.9115  & 30.880/0.9135  & 36.999/0.9699  & 30.710/0.9484  & 30.607/0.9430  & 31.542/0.9373  \\
        MEG-Net     \cite{zhang2021end}          & $\times$4    & 28.749/0.9160  & 31.103/0.9177  & 37.287/0.9716  & 30.674/0.9490  & 30.771/0.9453  & 31.717/0.9399  \\
        DistgSSR    \cite{wang2022disentangling} & $\times$4    & 28.992/0.9195  & 31.380/0.9217  & 37.563/0.9732  & 30.994/0.9519  & 31.649/0.9534  & 32.116/0.9439  \\
        LF-InterNet \cite{wang2020spatial}       & $\times$4    & 28.812/0.9162  & 30.961/0.9161  & 37.150/0.9716  & 30.777/0.9491  & 30.365/0.9409  & 31.613/0.9388  \\
        LF-IINet    \cite{liu2021intra}          & $\times$4    & 29.048/0.9188  & 31.331/0.9208  & 37.620/0.9734  & 31.039/0.9515  & 31.261/0.9502  & 32.060/0.9429  \\
        HLFSR-SSR   \cite{vinh2023-lfsr}         & $\times$4    & 29.196/0.9222  & 31.571/0.9238  & 37.776/0.9742  & 31.241/0.9543  & 31.641/0.9537  & 32.285/0.9456  \\
        LFSSR\_SAV  \cite{cheng2022spatial}      & $\times$4    & 29.368/0.9223  & 31.450/0.9217  & 37.497/0.9721  & 31.270/0.9531  & 31.362/0.9505  & 32.189/0.9439  \\
        DPT         \cite{wang2022detail}        & $\times$4    & 28.939/0.9170  & 31.196/0.9188  & 37.412/0.9721  & 30.964/0.9503  & 31.150/0.9488  & 31.932/0.9414  \\
        LFT         \cite{liang2022light}        & $\times$4    & 29.261/0.9209  & 31.433/0.9215  & 37.633/0.9735  & 31.218/0.9524  & 31.794/0.9543  & 32.268/0.9445  \\
        EPIT        \cite{Liang_2023_ICCV}       & $\times$4    & 29.339/0.9197  & 31.511/0.9231  & 37.677/0.9737  & 31.372/0.9526  & \underline{32.179}/0.9571  & 32.416/0.9452  \\
        LF-DET      \cite{cong2023lfdet}         & $\times$4    & 29.473/0.9230  & 31.558/0.9235  & 37.843/0.9744  & 31.388/0.9534  & 32.139/\underline{0.9573}  & 32.480/0.9463  \\
{\color{black}{MLFSR  \cite{Gao_2024_ACCV}}}       & {\color{black}{$\times$4}}    & {\color{black}{29.283/0.9218}}  & {\color{black}{31.564/0.9235}}  & {\color{black}{37.831/0.9745}}  & {\color{black}{31.241/0.9531}}  & {\color{black}{32.031/0.9567}} & {\color{black}{32.389/0.9235}} \\
{\color{black}{LFMamba  \cite{xia2024lfmamba}}}    & {\color{black}{$\times$4}}    & {\color{black}{\textbf{29.840}/\textbf{0.9256}}}  & {\color{black}{\textbf{31.695}/\textbf{0.9249}}}  & {\color{black}{\textbf{37.912}/\textbf{0.9748}}}  & {\color{black}{\textbf{31.808}/\textbf{0.9551}}}  & {\color{black}{31.846/0.9553}} & {\color{black}{\underline{32.620}/\textbf{0.9471}}} \\
        $L^2$FMamba (Ours)                       & $\times$4    & \underline{29.681}/\underline{0.9233}  & \underline{31.647}/\underline{0.9243}  & \underline{37.864}/\underline{0.9745}  & \underline{31.728}/\underline{0.9543}  & \textbf{32.198}/\textbf{0.9574}  & \textbf{32.623}/\underline{0.9468}  \\
        \bottomrule[2pt]
    \end{tabular}
    \label{tab:quantitative}
    \vspace{-5pt}
\end{table*}

\section{Experiments}

\subsection{Datasets and Implementation Details}
Based on previous studies \cite{jin2020deep, wang2020light, zhang2019residual}, we conducted experimental validation on five datasets, which include three real LF datasets: EPFL \cite{rerabek2016new}, INRIA \cite{le2018light}, and STF-gantry \cite{vaish2008new}, as well as two synthetic LF datasets:  HCIold \cite{wanner2013datasets} and HCInew \cite{honauer2017dataset}. We followed the same data protocol to partition the training and test sets.
We angularly cropped the central $5 \times 5$ SAIs to generate training and test samples.
During training, we cropped each SAI into square patches with sides of 64 / 128 and then applied $2 \times$ / $4 \times$ bicubic downsampling to generate LR patches, respectively.
We employed peak signal-to-noise ratio (PSNR) and structural similarity (SSIM) as evaluation metrics.
Firstly, we computed the average metric scores of the Y channel across all SAIs within each scene as the evaluation metric for that scene. 
Finally, we averaged the metric scores across all scenes in the dataset to obtain the evaluation metric for the dataset.

The default configuration of our $L^2$FMamba network is as follows: the initial feature dimension $C=64$, the number of LF-VSSM blocks $K=4$, the SSM state dimension $\text{d\_state}=16$, and the SSM feature expansion factor $\text{ssm-ratio}=1.0$, aiming to balance performance and efficiency. 
We employed the Adam optimizer with L1 loss and the StepLR scheduler, where the learning rate decreased by a factor of 0.5 every 30 epochs. The model was trained with an initial learning rate of $2.5 \times 10^{-4}$ for 180 epochs. 
We augmented the training data using random horizontal and vertical flipping, as well as 90-degree rotations.

\subsection{Comparison with the State-of-The-Arts}
We compared our method $L^2$FMamba to 16 SOTA methods, including both SISR and LFSR methods. 
Specifically, RCAN \cite{zhang2018image} is an SISR method based on CNN models.
{\color{black}{In the LFSR methods, except for DPT \cite{wang2022detail}, LFT \cite{liang2022light}, EPIT \cite{Liang_2023_ICCV}, and LF-DET \cite{cong2023lfdet}, which are based on the Transformer, as well as LFMamba \cite{xia2024lfmamba} and MLFSR \cite{Gao_2024_ACCV}, which are based on SSM, all other methods are based on the CNN models (\emph{i.e.}, LFSSR \cite{yeung2018light}, MEG-Net \cite{zhang2021end}, LFSSR\_SAV \cite{cheng2022spatial}, LF-ATO \cite{jin2020light}, LF-IINet \cite{liu2021intra}, DistgSSR \cite{wang2022disentangling}, and HLFSR-SSR \cite{vinh2023-lfsr}). }}
To achieve a fair comparison, the above methods were trained under the same experimental conditions.


\begin{enumerate}[wide]
\item{Quantitative Results:
The quantitative results of $L^2$FMamba and other SOTA methods are presented in Table \ref{tab:quantitative}.
{\color{black}{Our method demonstrates competitive results across all datasets for 2$\times$ and 4$\times$ LFSR tasks.
Specifically, in the 4$\times$ LFSR task, our method achieves the best results. Overall, our performance is comparable to LFMamba. However, on the STFgantry dataset, which features large disparities, our method outperforms it by 0.35 dB. 
This demonstrates that our strategy of directly modeling angular information on SAIs and MacPI is more effective in capturing long-range dependencies. 
In the 2$\times$ LFSR task, where the advantage of long-range modeling is relatively weakened, our method still achieves performance comparable to MLFSR. Notably, in subsequent experiments, we further validate the scalability of our method, showing that a significant performance improvement can be achieved with only a slight increase in parameters.
}}

}

\begin{figure*}[!t]
    \centering
    \vspace{-10pt}
    \includegraphics[width=18cm]{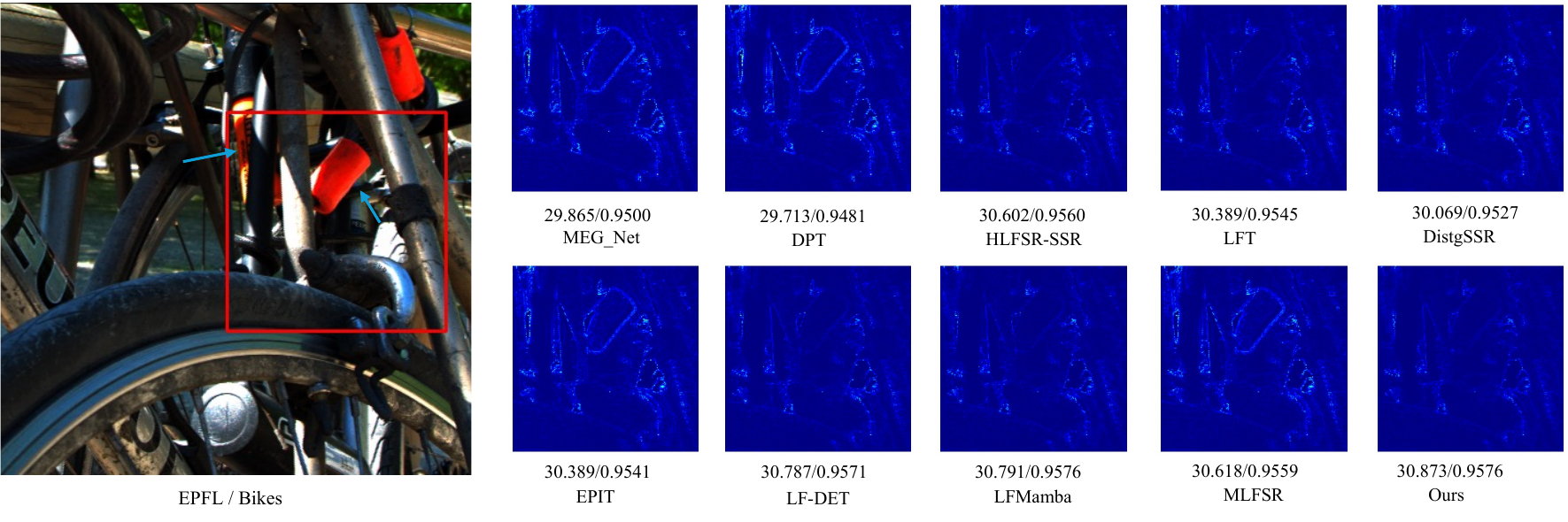}
    \quad
    \includegraphics[width=18cm]{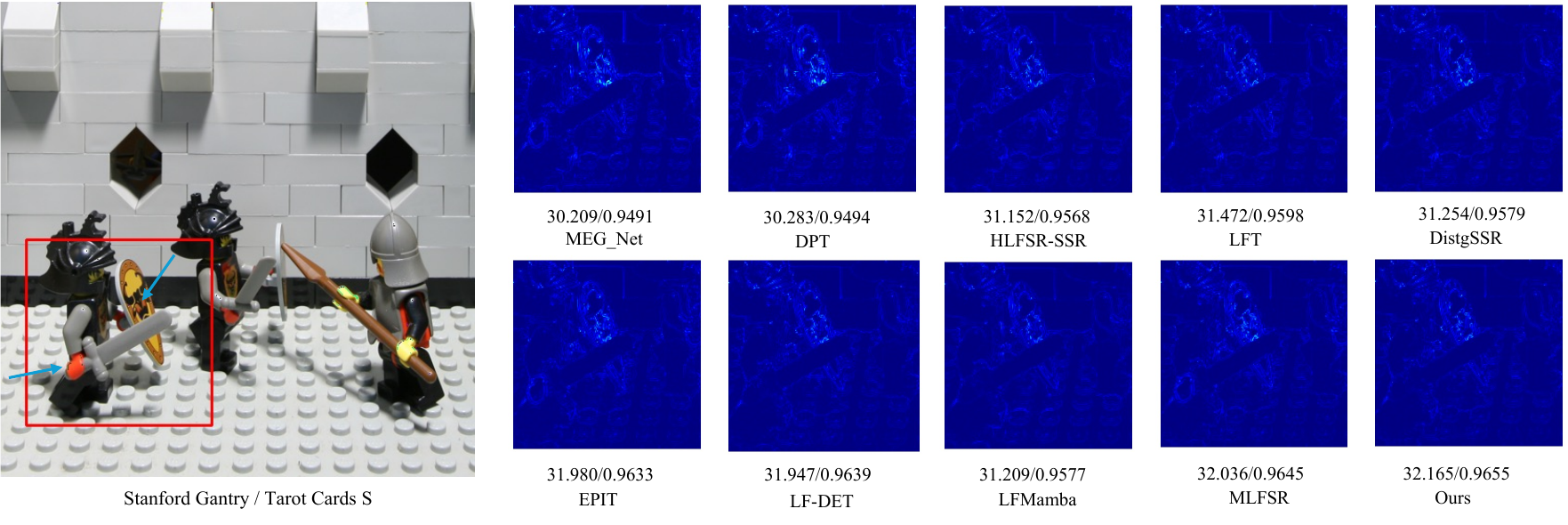}
    \caption{
            {\color{black}{Qualitative visualization results {\color{black}{for}} 4$\times$ LFSR compared to other methods. 
            Here, we show the error maps of the reconstructed {\color{black}{center-view}} images, with representative regions indicated by arrows.}}
            PSNR/SSIM values for the corresponding region are provided below.}
    \label{fig:4x}
\end{figure*}


\begin{table*}[!t]
    \fontsize{6.5pt}{8pt}\selectfont
    \centering
    \caption{Comparison of parameters, flops, time, and average PSNR/SSIM values for $\times$2 and $\times$4 SR. 
             Flops and time are calculated on an input LF with a size of $5\times5\times32\times32$.}
    \begin{minipage}[htp]{0.45\textwidth}
    \centering
    \begin{tabular}{@{}c@{\hspace{3pt}}c@{\hspace{4pt}}c@{\hspace{3pt}}c@{\hspace{3pt}}c@{\hspace{3pt}}c@{}}
        \toprule[2pt]
        Method                                  & Scale         & Params.     & FLOPs(G)     & Time(ms)       & {\color{black}{Avg. PSNR/SSIM}}  \\ 
        \midrule[1pt]  
        LFSSR      \cite{yeung2018light}        & $\times$2     & 0.89M       & 25.70        & 10.0           & 37.501/0.9832   \\
        MEG-Net    \cite{zhang2021end}          & $\times$2     & 1.69M       & 48.40        & 31.2           & 38.141/0.9851   \\
        LFSSR\_SAV \cite{cheng2022spatial}      & $\times$2     & 1.22M       & 31.11        & 14.0           & 38.262/0.9851   \\
        LF-ATO     \cite{jin2020light}          & $\times$2     & 1.22M       & 597.66       & 85.6           & 38.306/0.9847   \\
        LF-IINet   \cite{liu2021intra}          & $\times$2     & 5.04M       & 56.16        & 20.6           & 38.763/0.9860   \\
        DistgSSR   \cite{wang2022disentangling} & $\times$2     & 3.53M       & 64.11        & 24.2           & 38.940/0.9867   \\
        HLFSR-SSR  \cite{vinh2023-lfsr}         & $\times$2     & 13.72M      & 167.81       & 31.1           & 39.303/0.9874   \\
        LFT        \cite{liang2022light}        & $\times$2     & 1.11M       & 56.16        & 91.4           & 38.825/0.9861   \\
        DPT        \cite{wang2022detail}        & $\times$2     & 3.73M       & 65.34        & 98.5           & 38.397/0.9848   \\
        EPIT       \cite{Liang_2023_ICCV}       & $\times$2     & 1.42M       & 69.71        & 32.2           & 39.393/0.9869   \\
        LF-DET     \cite{cong2023lfdet}         & $\times$2     & 1.59M       & 48.50        & 65.9           & 39.455/0.9874   \\
        {\color{black}{MLFSR  \cite{Gao_2024_ACCV}}}       & {\color{black}{$\times$2}}    & {\color{black}{1.36M}}  & {\color{black}{53.30}}  & {\color{black}{27.8}}  & {\color{black}{39.231/0.9873}} \\
        {\color{black}{LFMamba \cite{xia2024lfmamba}}}     & {\color{black}{$\times$2}}    & {\color{black}{2.15M}}  & {\color{black}{92.29}}  & {\color{black}{75.9}}  & {\color{black}{39.424/0.9881}} \\
        $L^2$FMamba (Ours)                      & $\times$2     & 1.04M       & 36.59        & 31.9           & 39.485/0.9873   \\
        \midrule[1pt]  
        \end{tabular}
        \end{minipage}%
        \begin{minipage}[htp]{0.45\textwidth}
        \centering
        \begin{tabular}
{@{}c@{\hspace{3pt}}c@{\hspace{4pt}}c@{\hspace{3pt}}c@{\hspace{3pt}}c@{\hspace{3pt}}c@{}}
        \toprule[2pt]
        Method                                  & Scale         & Params.     & FLOPs(G)     & Time(ms)       & {\color{black}{Avg. PSNR/SSIM}}  \\ 
        \midrule[1pt]  
        LFSSR      \cite{yeung2018light}        & $\times$4     & 1.61M       & 128.44       & 37.7           & 31.517/0.9370   \\
        MEG-Net    \cite{zhang2021end}          & $\times$4     & 1.77M       & 102.20       & 32.4           & 31.717/0.9399   \\
        LFSSR\_SAV \cite{cheng2022spatial}      & $\times$4     & 1.54M       &  99.15       & 43.2           & 32.189/0.9439   \\
        LF-ATO     \cite{jin2020light}         & $\times$4     & 1.66M       & 686.99       & 88.7           & 31.542/0.9373   \\
        LF-IINet   \cite{liu2021intra}         & $\times$4     & 4.89M       & 57.42        & 20.8           & 32.060/0.9429   \\
        DistgSSR   \cite{wang2022disentangling} & $\times$4     & 3.58M       & 65.41        & 25.0           & 32.116/0.9439   \\
        HLFSR-SSR  \cite{vinh2023-lfsr}         & $\times$4     & 13.87M      & 182.93       & 32.7           & 32.285/0.9456  \\
        LFT        \cite{liang2022light}       & $\times$4     & 1.16M       & 57.60        & 95.2           & 32.268/0.9445   \\
        DPT        \cite{wang2022detail}        & $\times$4     & 3.78M       & 66.55        & 99.7           & 31.932/0.9414   \\
        EPIT       \cite{Liang_2023_ICCV}       & $\times$4     & 1.47M       & 71.15        & 33.6           & 32.416/0.9452   \\
        LF-DET     \cite{cong2023lfdet}         & $\times$4     & 1.69M       & 51.20        & 75.0           & 32.480/0.9463   \\
        {\color{black}{MLFSR  \cite{Gao_2024_ACCV}}}       & {\color{black}{$\times$4}}    & {\color{black}{1.41M}}  & {\color{black}{54.74}}  & {\color{black}{28.9}}  & {\color{black}{32.389/0.9235}} \\
        {\color{black}{LFMamba  \cite{xia2024lfmamba}}}    & {\color{black}{$\times$4}}    & {\color{black}{2.30M}}  & {\color{black}{96.24}}  & {\color{black}{77.1}}  & {\color{black}{32.620/0.9471}} \\
        $L^2$FMamba (Ours)                     & $\times$4     & 1.09M       & 37.99        & 32.6           & 32.623/0.9468   \\
        \bottomrule[1pt]
    \end{tabular}
    \end{minipage}%
    \label{tab:efficiency}
\end{table*}

\begin{table*}
    \centering
    \caption{Ablation study on different components for 4$\times$ LFSR.}
    \begin{tabular}{c@{\hspace{20pt}}ccc@{\hspace{20pt}}cc@{\hspace{20pt}}c}
        \toprule[2pt]
        Index & First & Second & Thrid & Params.   & FLOPs(G)     & {\color{black}{Avg. PSNR/SSIM}}  \\ 
        \midrule[1pt]  
        a)    & {\color{black}{Intra-SSM}}   & {\color{black}{Intra-SSM}}   & {\color{black}{Intra-SSM}}   & 1.094M & 38.105 & 30.407/0.9147 \\ 
        b)    & {\color{black}{Inter-SSM}}  & {\color{black}{Inter-SSM}}  & {\color{black}{Inter-SSM}}  & 1.086M & 37.928 & 31.196/0.9334 \\ 
        c)    & MacPI-SSM & MacPI-SSM & MacPI-SSM & 1.086M & 37.928 & 32.546/0.9462 \\
        \midrule[1pt]
        d)    & {\color{black}{Intra-SSM}}   & {\color{black}{Intra-SSM}}   & MacPI-SSM & 1.091M & 38.046 & 32.503/0.9462 \\
        e)    & {\color{black}{Intra-SSM}}   & MacPI-SSM & MacPI-SSM & 1.088M & 37.987 & 32.569/0.9463 \\
        f)    & MacPI-SSM & {\color{black}{Inter-SSM}}  & {\color{black}{Intra-SSM}}   & 1.088M & 37.987 & 32.490/0.9462 \\
        \midrule[1pt]
        Our   & {\color{black}{Intra-SSM}}   & {\color{black}{Inter-SSM}}  & MacPI-SSM & 1.088M & 37.987 & 32.623/0.9468 \\
        \toprule[2pt]
    \end{tabular}
   
    \label{tab:component_and_order}
\end{table*}

\item{\emph{Qualitative Results:}
{\color{black}{Fig. \ref{fig:4x} presents the qualitative results of error maps for different methods on the 4$\times$ LFSR task. 
Compared to other methods, our approach demonstrates superior texture detail restoration and edge sharpness. 
In different scenes, such as the bottle contours and text details in "EPFL/Bikes", and the hand contours and shield details in "Stanford Gantry/Tarot Cards S", our method shows outstanding performance. 
Our model not only leads in detail preservation but also achieves the highest PSNR and SSIM scores in the corresponding regions, demonstrating excellent performance in both {\color{black}{quantitative and qualitative}}  evaluations. In particular, compared to LFMamba, our method achieves similar or even superior qualitative and quantitative results, with fewer parameters and lower complexity, which fully demonstrates the efficiency of our approach.}}}

\item{\emph{Computational Efficiency:}
We comprehensively compared the computational efficiency of $L^2$FMamba with other methods in terms of parameters, FLOPs, and inference time.
As shown in Table \ref{tab:efficiency}, the proposed method achieves SOTA performance in the LFSR task with fewer model parameters, lower computational complexity, and faster inference. 
{\color{black}{It is worth noting that our method currently lags behind typical CNN-based methods (such as LFSSR, LFSSR-SAV, LF-IINet, and DistgSSR) in terms of inference speed. 
This is mainly due to CNN architectures benefiting from highly optimized computational libraries, such as cuDNN, which fully leverage the parallel computing power of GPUs. 
In contrast, our method is based on the SSM module, while theoretically excelling in modeling long-range dependencies, still faces ongoing engineering optimization and hardware acceleration support. 
Nevertheless, compared to Transformer-based methods such as EPIT, HLFSR-SSR, and LF-DET, our method reduces the model parameter count by 35\%-48\%, lowers computational complexity by 25\% compared to LF-DET, and improves inference speed by 40\%-50\%. 
These comparative experiments clearly demonstrate that $L^2$FMamba effectively addresses the inherent challenges of high computational complexity and memory consumption in Transformer architectures, while maintaining excellent performance in super-resolution.}}

}
\end{enumerate}


\subsection{Ablation Study}

\begin{enumerate}[wide]

\item{Effectiveness of Components: 
This section explores the effectiveness of different layers within LF-VSSM. We conducted three sets of ablation experiments, each retaining only one specific layer. To prevent the number of parameters from influencing the results, we maintained a consistent 3-layer structure in each block. The specific configurations are shown in rows a), b), and c) of Table \ref{tab:component_and_order}. 

\emph{{\color{black}{Intra-SSM}} Layer:} When only the {\color{black}{Intra-SSM}} layer is used, it primarily extracts spatial information from LF images while completely ignoring angular information, effectively reducing the task to a standard SISR problem. The results in row a) are similar to those of RCAN, further validating the spatial feature extraction capability of the {\color{black}{Intra-SSM}} layer.

\emph{{\color{black}{Inter-SSM}} Layer:} The results in row b) show a slight improvement over row a), indicating the utilization of the angular information between SAIs. However, the focus remains on spatial information extraction, aligning with our design approach. Thus, while the {\color{black}{Inter-SSM}} layer contributes to angular information extraction, it does not fully exploit it without the support of other layers.

\emph{MacPI-SSM Layer:} The results of using only the MacPI-SSM layer in row c) show a significant improvement compared to rows a) and b), even surpassing advanced methods like LF-DET in terms of effectiveness. Thus, the MacPI-SSM layer can effectively extract long-range spatial-angular contextual dependencies from light field images. However, there is still a 0.1 dB gap compared to our proposed LF-VSSM, further validates the effectiveness of our progressive spatial-angular feature extraction approach.
}

\item{Effectiveness of progressive features extraction: 
To validate the effectiveness of the proposed progressive feature extraction method, three ablation experiments, as shown in rows d), e), and f) of Table \ref{tab:component_and_order}, are designed. 
In experiments d) and e), we removed the {\color{black}{Inter-SSM}} layer and replaced it with {\color{black}{Intra-SSM}} and MacPI-SSM, respectively, to assess the contribution of angular information between SAIs to performance. 
In experiment f), we altered the order of progressive feature extraction to examine the impact of the order on the results. 
The results show that the performance of these three experiments is inferior to the original method, demonstrating the critical importance of progressive spatial-angular feature extraction for the LFSR task.
Additionally, the results of experiment e) are superior to those of experiment c), further validating the benefits of the progressive spatial-angular feature extraction strategy.}


\begin{figure*}[!t]

    \centering
    \subfigure[1st Epoch]{
        \includegraphics[width=18cm]{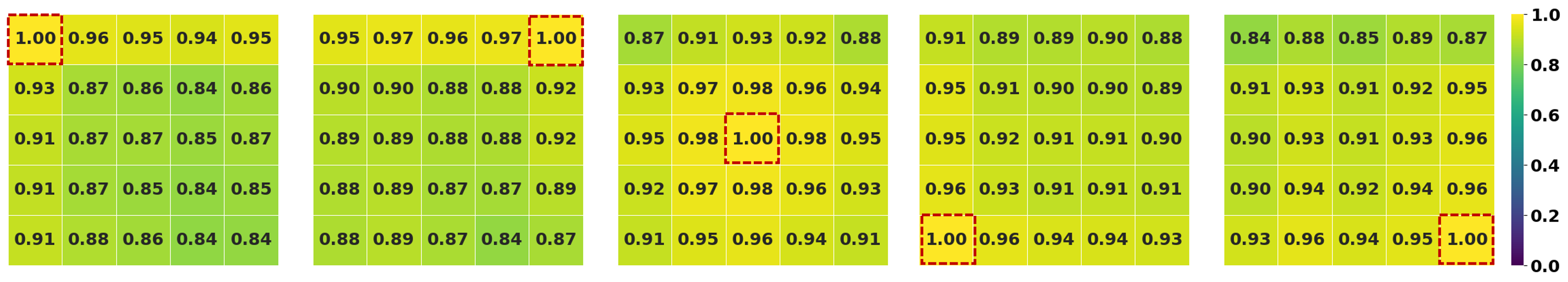}
    }
    \subfigure[30th Epoch]{
        \includegraphics[width=18cm]{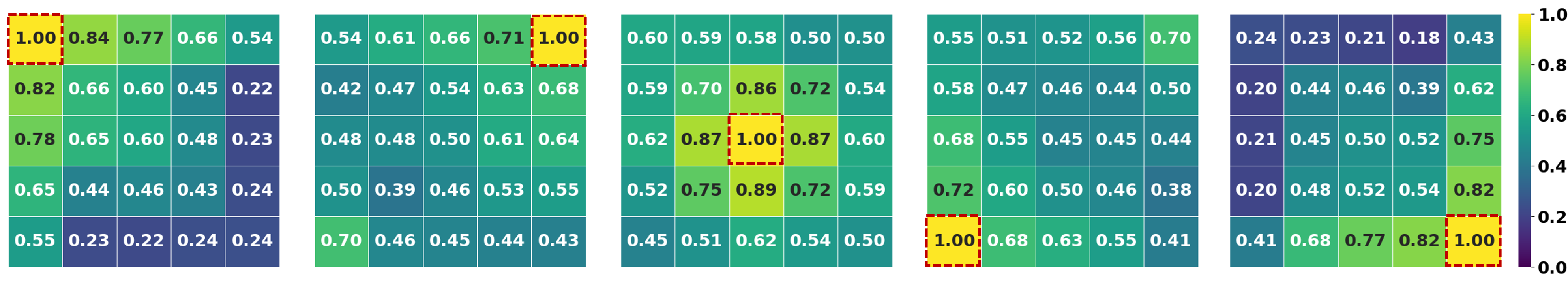}
    }
    \subfigure[60th Epoch]{
        \includegraphics[width=18cm]{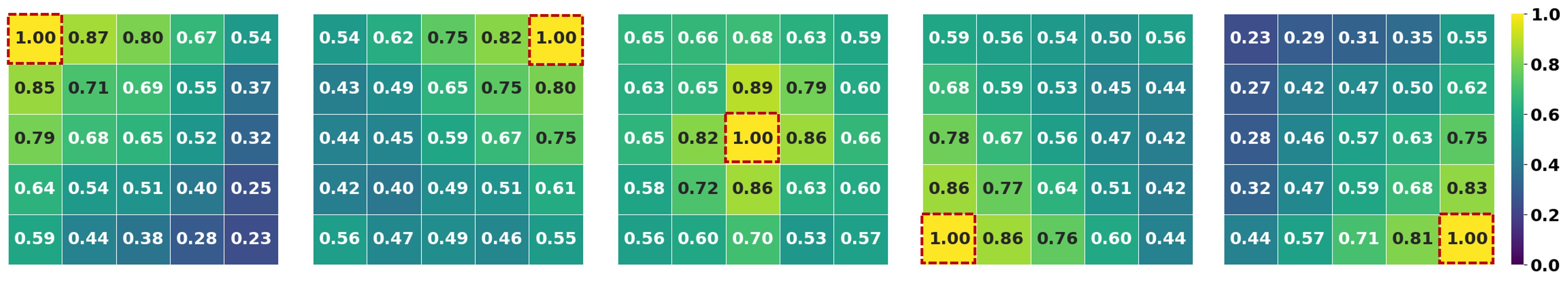}
    }
\caption{{\color{black}{Heatmap of similarity between angular embeddings at top-left, bottom-left, center, bottom-right, top-right positions and other angular embeddings across different training epochs. 
                       Red boxes indicate anchors in the heatmap, Numbers represent the similarity between anchors and angular embeddings.}}}
\label{fig:angpos}
\vspace{-10pt}
\end{figure*}

\begin{table}[t]
    \centering
    \caption{Ablation study on different D\_STATE (D) and SSM\-RATIO (R) for 4$\times$ LFSR.}
    \begin{tabular}{@{}cccccc@{}}
        \toprule[2pt]
        D & R & Params.   & FLOPs(G)   & {\color{black}{Avg. PSNR/SSIM}}  \\ 
        \midrule[1pt]  
         8 & 1.0 & 1.015M & 31.065 & 32.438/0.9452 \\
         8 & 2.0 & 1.271M & 42.627 & 32.526/0.9461 \\
        16 & 1.0 & 1.088M & 37.987 & 32.623/0.9468 \\
        16 & 2.0 & 1.490M & 56.599 & 32.720/0.9476 \\
        \toprule[2pt]
    \end{tabular}
    \vspace{-10pt}
    \label{tab:parameters}
\end{table}

\begin{table}[t]
    \centering
    \caption{Ablation study {\color{black}{on}} position embedding for 4$\times$ LFSR.}
    \begin{tabular}{@{}ccccc@{}}
        \toprule[2pt]
                       & Params.   & FLOPs(G)   & {\color{black}{Avg. PSNR/SSIM}}  \\ 
        \midrule[1pt]  
        w/o $P_{ang}$  & 1.087M  & 37.987  & 32.503/0.9457 \\ \\
        {\color{black}{w/ $P_{spa}^{fix}$}}     & 1.154M  & 37.987  & 32.608/0.9467 \\ \\
        {\color{black}{w/ $P_{spa}^{able}$}} & 1.154M  & 37.987  & 32.615/0.9468 \\ \\
        $L^2$FMamba    & 1.088M  & 37.987  & 32.623/0.9468 \\
        \toprule[2pt]
    \end{tabular}
   
    \label{tab:position}
\end{table}

\begin{table}[t]
    \centering
    \caption{Ablation study on different LF-VSSM blocks (K) for 4$\times$ LFSR.}
    \begin{tabular}{@{}ccccc@{}}
        \toprule[2pt]
        K & Params.   & FLOPs(G)   & {\color{black}{Avg. PSNR/SSIM}}  \\ 
        \midrule[1pt]  
        2 & 0.634M  & 21.373G & 32.017/0.9419 \\
        4 & 1.088M  & 37.987G & 32.623/0.9468 \\
        6 & 1.543M  & 54.600G & 32.700/0.9476 \\
        8 & 1.998M  & 71.214G & 32.754/0.9483 \\
        \toprule[2pt]
    \end{tabular}
    \label{tab:num_blocks}
    \vspace{-10pt}
\end{table}

\item{Position Embedding:}
{\color{black}{We conducted three sets of ablation experiments related to positional encoding to explore its impact on model performance, as shown in Table \ref{tab:position}. In the $w/o\ P_{ang}$ experiments, we removed the angular positional encoding, which resulted in a significant performance drop ($- 0.121$dB), clearly demonstrating the importance of angular positional encoding. 
Building on the addition of angular positional encoding, we further introduced two types of spatial positional embeddings: $w/\ P_{spa}^{fix}$ with fixed sine/cosine positional encoding and $w/\ P_{spa}^{able}$ with learnable spatial positional encoding. We found that $w/\ P_{spa}^{able}$ performed slightly better than $w/\ P_{spa}^{fix}$, but the impact on the final results was minimal.
This finding supports the conclusion that the scanning path design in SSM already provides sufficient spatial positional information.}}

{\color{black}{Additionally, we conducted a visual analysis {\color{black}{on}} the $P_{ang}$ during training. 
We specifically selected the 1st, 30th, and 60th epochs, as they effectively capture the evolution of the $P_{ang}$ throughout the training process. 
The 1st epoch shows the initial random state, while the 30th epoch reflects the early increase in encoding similarity. 
After the 60th epoch, we observe that the changes of the $P_{ang}$ stabilize, indicating that the model has learned a stable geometric structure, with no significant changes thereafter.
We computed the cosine similarity between the angular embeddings at the upper-left, upper-right, center, lower-left, and lower-right positions and those at other locations, and visualized the results as heatmaps, as shown in Fig. \ref{fig:angpos}. 
From the heatmap, it is evident that as training progresses, the similarity between $P_{ang}$ increases, gradually aligning with the geometric structure of the LF angular distribution. 
This progression demonstrates that the model has successfully captured the spatial relationships between different viewpoints. 
This transformation provides robust prior knowledge for subsequent feature extraction and further highlights the crucial role of angular positional encoding in enhancing model performance.}}



\item{SSM Parameters:}
The d\_state and ssm-ratio are critical parameters in SSM.
We conducted a series of ablation experiments to explore their impact on $L^2$FMamba, with the results presented in Table \ref{tab:parameters}. 
The experimental results indicate that d\_state has a more significant influence on LFSR task performance, likely because a higher d\_state enables the model to capture and represent more complex dynamic information. 
Conversely, ssm-ratio primarily affects the parameter count and computational complexity. To achieve an optimal balance between performance and efficiency, we set the d\_state to 8 and the ssm-ratio to 1.0 as the default configuration for $L^2$FMamba.

\item{Number of LF-VSSM Blocks: 
Finally, we investigated the impact of the number of LF-VSSM blocks on LFSR task performance, as shown in Table \ref{tab:num_blocks}.
The results demonstrate that $L^2$FMamba can enhance expressive capability and performance by increasing the number of LF-VSSM blocks, demonstrating its good scalability. 
However, as the number of blocks increases, the computational parameters and complexity also rise significantly. Therefore, we ultimately select 4 LF-VSSM blocks in $L^2$FMamba.}
\end{enumerate}

\section{Conclusion}
This study first introduces the LF-VSSM block for the LFSR task.
Based on the principle of progressive feature extraction, it successively captures critical long-range spatial-angular context dependencies within LF images across spatial and angular dimensions, significantly enhancing the performance of LFSR tasks.
On this basis, we propose a lightweight network, $L^2$FMamba, which integrates the LF-VSSM block. $L^2$FMamba effectively reduces the number of parameters and computational costs while achieving {\color{black}{state-of-the-art (SOTA)}} performance.
Extensive experimental results on five common LF datasets demonstrate that our method achieves SOTA results for 2$\times$ and 4$\times$ LFSR tasks, further validating the effectiveness, efficiency, and scalability of the proposed network.

However, despite the significant progress made by $L^2$FMamba in the LFSR task, there are still some limitations. 
The method shows insufficient detail recovery when handling large disparity variations, and the reconstruction of fine structures at high magnification is suboptimal. 
Future research could enhance robustness and recovery capability by incorporating multi-level feature fusion and optimizing the structure of the lightweight network.
\bibliographystyle{IEEEtran}
\bibColoredItems{black}{Gao_2024_ACCV}
\bibColoredItems{black}{hu2024beyond}
\bibliography{IEEEabrv,ref}











\newpage

 





\end{document}